\documentclass[aps,pra,twocolumn]{revtex4}
\usepackage{longtable}
\usepackage{dcolumn}
\newcolumntype{.}{D{x}{}{-1}}

\usepackage[dvips]{graphicx}
\usepackage{bm}
\usepackage{bbm}

\usepackage{times}
\usepackage{nicefrac}
\usepackage{amsmath}
\usepackage{amsfonts}
\usepackage{amssymb}
\usepackage{amsthm}

\newcolumntype{.}{D{x}{}{-1}}

\newcommand{\vare}{\varepsilon}

\newcommand{\lbr}{\langle}
\newcommand{\rbr}{\rangle}

\newcommand{\Za}{Z\alpha}

\begin{document}

\title{Relativistic
configuration-interaction calculations of energy levels of the $\bm{1s^22l}$ and $\bm{1s2l2l'}$
states in lithium-like ions: carbon through chlorine}

\author{V.~A. Yerokhin}
\affiliation{Physikalisch-Technische Bundesanstalt, D-38116 Braunschweig, Germany}
\affiliation{Center for Advanced Studies, Peter the Great St.~Petersburg Polytechnic University,
195251 St.~Petersburg, Russia}

\author{A.~Surzhykov}
\affiliation{Physikalisch-Technische Bundesanstalt, D-38116 Braunschweig, Germany}
\affiliation{Technische Universit\"at Braunschweig, D-38106 Braunschweig, Germany}

\author{A.~M\"uller}
\affiliation{Institut f\"ur Atom- und Molek\"ulphysik, Justus-Liebig-Universit\"at Gie{\ss}en,
D-35392 Giessen, Germany}

\begin{abstract}

We present systematic calculations of energy levels of the $1s^22l$ and $1s2l2l'$ states of ions
along the lithium isoelectronic sequence from carbon till chlorine. The calculations are
performed by using the relativistic-configuration-interaction method adapted to the treatment of
autoionizing core-excited states. The relativistic energies are supplemented with the QED energy
shifts calculated within the model QED operator approach. A systematic estimation of the
theoretical uncertainties is performed for every electronic state and every nuclear charge. The
results are in agreement with existing high-precision theoretical and experimental data for the
ground and first excited states. For the core-excited states, our theory is much more accurate
than the presently available measurements.

\end{abstract}

\maketitle

\section{Introduction}

Lithium-like ions are among the simplest atomic systems. Their spectra can be described in {\em ab
initio} theoretical calculations very accurately. For light atoms, the presently most powerful
calculational approach is based on the nonrelativistic quantum electrodynamics (NRQED) expansion of
 energy levels in powers of $\alpha$ and $Z\alpha$ (where $\alpha$ is the fine-structure
constant and $Z$ is the nuclear charge number). High-precision NRQED calculations were performed by
Puchalski and Pachucki for the lowest-lying states of Li and Be$^+$
\cite{puchalski:08:li,puchalski:14,puchalski:15}. In the region of heavy ions, the best results are
presently obtained within the alternative approach that accounts for all orders in the nuclear
binding strength parameter $Z\alpha$ but expands in the electron-electron interaction parameter
$1/Z$. Calculations by this method were performed by Shabaev and co-workers
\cite{yerokhin:01:2ph,yerokhin:06:prl,yerokhin:07:lilike,kozhedub:10} for the ground and first
excited states of Li-like ions with $Z \ge 10$. It is important to point out that both these
methods were able to produce {\em predictive} results, i.e.,  their results contain estimates of
theoretical errors obtained without referring to experimental data.

For the core-excited states of Li-like ions, there have been no rigourous QED calculations
accomplished so far. Previous calculations were performed using various methods, notably, the
multi-configurational Dirac-Fock method \cite{chen:86,nilsen:88}, the variational nonrelativistic
approach with inclusion of leading relativistic effects \cite{chung:84}, and many-body perturbation
theory (MBPT) \cite{safronova:04:cjp}. None of these calculations were able to provide estimations
of theoretical errors.

In our previous investigation \cite{yerokhin:12:lilike} we obtained predictive results for energies
of the $1s2l2l'$ core-excited states of Li-like ions in the nuclear charge region $Z = 18 - 36$. By
combining results obtained by the relativistic configuration-interaction method with the one-loop
QED effects calculated in effective screening potentials, we were able to produce theoretical
predictions with an accuracy better than what is presently achievable in experiments, see, e.g.,
the recent measurement of the $K\alpha$ transitions in iron \cite{rudolph:13}. Such accuracy opens
possibilities of using theoretical energies of Li-like ions for calibration of experimental X-ray
spectra for ions with a larger number of electrons, for which accurate calculations are presently
not possible.

With the range of $Z$ computed in Ref.~\cite{yerokhin:12:lilike},  possible calibrations are
restricted to the X-ray energies beyond 3~keV. The energy range of most third- and
fourth-generation synchrotron light sources, however, lies in the region of smaller X-ray energies
\cite{couprie:15}. High-quality calibration sources are urgently needed for the energy range
between the carbon and the chlorine $K$-edges.

The present situation with calibration standards in the soft-X-ray regime  has recently been
examined by M\"{u}ller and coworkers~\cite{mueller:17:neon} with emphasis on photon energies near
the neon $K$-edge at approximately 870~eV. In numerous experiments performed by various techniques
(see Table~I of Ref.~\cite{mueller:17:neon}), the $1s$$\to$$3p$ dipole transition energy in neutral
neon has been measured, with results ranging between 867.05 and 867.69~eV with quoted uncertainties
of typically 50 to 80~meV but  discrepancies reaching up to 640~meV. This situation clearly shows
the need for new and reliable calibration standards in the soft-X-ray energy region.

Precise knowledge of the satellite transition energies are also required for the diagnostics of hot
laboratory plasmas, particularly those in the magnetically-confined-fusion research. High-quality
theoretical energies are considered to be critical for a proper fit of spectral lines and,
accordingly, a better plasma diagnostics \cite{ralchenko:priv,piron:17}.

Partly motivated by the needs described above, the goal of the present work was to extend our
previous calculation  of the $n = 2$ valence and core-excited states of Li-like ions
\cite{yerokhin:12:lilike} to the lower-$Z$ region. This task turned out to be less straightforward
than it seemed and required significant alterations of our original computational approach, for two
reasons. First, the interaction of the autoionizing core-excited reference states with
closely-lying continuum states became more pronounced for low-$Z$ ions than it was for heavier
ions, which led to poor convergence of the results with respect to the basis size. Second, the
computation of the QED effects for the nuclear charges as low as $Z = 6$ in the same way as it was
done in Ref.~\cite{yerokhin:12:lilike} turned out to be not possible because of technical
difficulties and numerical cancellations, which grow fast as $Z$ is decreased. Our ways for
overcoming these problems are discussed in the next two sections.

\section{Configuration-Interaction method for core excited states}

\begin{table*}
\caption{The convergence of the CI energies for the
  $1s2p^2\,^2\!D_{3/2}$ state (in a.u.) with respect to the number of one-electron orbitals in the basis,
  for the standard and for the balanced $B$-spline basis (see
  text). For illustration
  purposes, the one-electron basis is restricted to contain orbitals with $l\le 2$ only and
  only the Coulomb interaction is included into the Hamiltonian. $n_a$ is the number of
  $B$ splines, $N_{\rm orb}$ is the number of
  one-electron orbitals, $E$ is the energy value. \label{tab:checkgridCI}}
\begin{ruledtabular}
\begin{tabular}{lr..r..}
          & \multicolumn{3}{c}{Standard basis}
                                 & \multicolumn{3}{c}{Balanced basis}   \\
 \multicolumn{1}{l}{$n_a$}
          & $N_{\rm orb}$  & \multicolumn{1}{c}{$E$} & \multicolumn{1}{c}{Increment}
          & $N_{\rm orb}$  & \multicolumn{1}{c}{$E$} & \multicolumn{1}{c}{Increment}
\\
\hline\\[-5pt]
%
 \multicolumn{1}{l}{$Z = 6$}\\

30 &   81 &   -23.x536\,9  &             &   86 &    -23.x519\,37      &                    \\
40 &   98 &   -23.x519\,1  &  0.x017\,8  &  101 &    -23.x519\,31      & 	 0.x000\,06     \\
50 &  118 &   -23.x523\,4  & -0.x004\,4  &  113 &    -23.x519\,28      &	 0.x000\,03     \\	
60 &  141 &   -23.x514\,7  &  0.x008\,7  &  148 &    -23.x519\,25      & 	 0.x000\,03     \\
70 &  161 &   -23.x517\,8  & -0.x003\,1  &  166 &    -23.x519\,24      &	 0.x000\,01     \\
80 &  178 &   -23.x517\,4  &  0.x000\,4  &  183 &    -23.x519\,23      &	 0.x000\,01     \\
Result&&      -23.x517\,(15)&            &      &    -23.x519\,23\,(5) & \\[5pt]
 \multicolumn{1}{l}{$Z = 17$}\\
30 &   81 &  -206.x754\,8  &             &   81 &   -206.x754\,26      &                  \\
40 &  101 &  -206.x753\,0  &  0.x001\,8  &  102 &   -206.x754\,11      & 	0.x000\,15     \\
50 &  118 &  -206.x759\,0  & -0.x006\,0  &  116 &   -206.x754\,05      &	0.x000\,06     \\	
60 &  146 &  -206.x749\,5  &  0.x009\,5  &  138 &   -206.x754\,01      & 	0.x000\,04     \\
70 &  163 &  -206.x752\,7  & -0.x003\,3  &  153 &   -206.x753\,98      &	0.x000\,03     \\
80 &  181 &  -206.x752\,7  &  0.x000\,0  &  170 &   -206.x753\,96      &	0.x000\,02     \\
Result&&     -206.x752\,(15) &           &      &   -206.x754\,0\,(2) \\[5pt]
\end{tabular}
\end{ruledtabular}
\end{table*}

We start with outlining the main features of the configuration-interaction (CI) method, which is by
now one of the standard approaches in atomic structure calculations, see, e.g.,
Refs.~\cite{chen:93:pra,chen:95}. The CI $N$-electron wave function $\Psi(PJM)$ with a definite
parity $P$, total angular momentum $J$, and angular momentum projection $M$ is represented as a
finite sum of configuration-state functions (CSFs) with the same $P$, $J$, and $M$,
\begin{equation}\label{eq4}
  \Psi(PJM) = \sum_r c_r \Phi(\gamma_r PJM)\,,
\end{equation}
where $\gamma_r$ denotes the set of additional quantum numbers that determine the CSF. The CSFs are
constructed as linear combinations of antisymmetrized products of one-electron orbitals $\psi_n$,
which are {\em positive-energy} eigenfunctions of some one-particle Dirac Hamiltonian (which
corresponds to the so-called no-pair approximation). In our implementation of the CI method, we
used the one-particle Dirac Hamiltonian with the frozen-core Dirac-Fock potential.

The eigenvalues and eigenfunctions of the Dirac Hamiltonian are constructed by the
dual-kinetic-balance (DKB) method \cite{shabaev:04:DKB} from a finite set of $B$-spline basis
functions. This approach yields a discrete representation of the continuum part of the Dirac
spectrum, in which the density of the continuum states increases as the number of basis functions
is enlarged. For a given number of B-splines $n_a$, all Dirac eigenstates $\psi_n$ with the
energies $0< \vare_n \le mc^2(1+Z\alpha\, \epsilon)$ and the orbital quantum number $l \le L_{\rm
max}$ were included into the one-electron basis of our CI calculations. The dependence of the
calculated results on the parameters $n_a$, $\epsilon$, and $L_{\rm max}$ was carefully studied in
order to provide estimates of the numerical uncertainty (see Tables I and II of
Ref.~\cite{yerokhin:12:lilike} for examples of the analysis of the basis convergence).

The energies of electronic states and the corresponding expansion coefficients $c_r$ are obtained
as the eigenvalues and the eigenvectors of the matrix of the Dirac-Coulomb-Breit (DCB) Hamiltonian
in the space of the CSFs,
\begin{eqnarray}\label{eq:0}
   \left\{ H_{rs}\right\} \equiv \left\{ \lbr \gamma_r PJM|H_{\rm DCB}|\gamma_s PJM\rbr\right\}\,.
\end{eqnarray}
The DCB Hamiltonian is
\begin{eqnarray}
    H_{\rm DCB} = \sum_i h_{\rm D}(i) + \sum_{i<j} \left[ V_{C}(i,j)+
    V_{B}(i,j)\right]\,,
\end{eqnarray}
where the indices $i,j = 1,\ldots,N$ numerate the electrons, $h_D$ is the one-particle
Dirac-Coulomb Hamiltonian, and $V_{C}$ and $V_B$ are the Coulomb and the Breit parts of the
electron-electron interaction. The matrix elements of the Hamiltonian are represented as linear
combinations of one- and two-particle radial integrals,
\begin{align}\label{eq7}
\lbr \gamma_r PJM| &\,  H_{\rm DCB}|\gamma_s PJM\rbr = \sum_{ab}
 d_{rs}(ab)\,I(ab)
  \nonumber \\
 + &\, \alpha \sum_k \sum_{abcd} v_{rs}^{(k)}(abcd)\,
  R_{k}(abcd)\,,
\end{align}
where $a$, $b$, $c$, and $d$ numerate the one-electron orbitals, $d_{rs}$ and $v^{(k)}_{rs}$ are
the angular coefficients, $I(ab)$ are the one-electron radial integrals, and $R_{k}(abcd)$ are the
two-electron radial integrals. We refer the reader to our previous papers for formulas and details
of the implementation of the method \cite{yerokhin:12:lilike,yerokhin:14:belike}.

In the present work we would like to use the CI method for computing energy levels of core-excited
states with energies above the autoionization threshold. For such states, the interaction of the
reference state with the closely-lying continuum $1s^2\vare l$ states (with energy $\vare > mc^2$)
might be significant and should be properly accounted for. In our previous calculations
\cite{yerokhin:12:lilike,yerokhin:14:belike,yerokhin:15:ps}, we addressed this issue by using
increasingly large sets of one-electron orbitals and by studying the convergence of the results. In
the present work, however, we are interested in the lower-$Z$ ions where the interaction with the
continuum states is more significant and the convergence of the results for such a straightforward
approach often becomes unsatisfactory.

Let us consider the $1s2s^2\,^2\!S$ state as an example. It is the lowest-lying core-excited state
and its energy is significantly influenced by the interaction with the closely-lying continuum
$1s^2 \vare p_{1/2}$ states. If we increase our basis of one-electron orbitals in a straightforward
way, we observe that at a certain level of precision our results  stop to converge but start to
oscillate instead. We would like to emphasize that this problem reveals itself only when we
increase the {\em density} of the continuum states in the ``problematic'' region near the energy of
the reference state. If we had not changed the density, we might have not even become aware of this
problem.

It turns out that the instabilities of the convergence of energies can be traced back to the
situations when a continuum $1s^2 \vare p_{1/2}$ state happens to be closely degenerate in energy
with the reference state. We found out that the convergence with respect to the basis can be
drastically improved if we ``balance'' our discrete representation of the continuum spectrum in
such a way that the energies of the two nearest continuum states are on the same distance from the
energy of the reference state (one continuum state below the reference state, and the other above).

The same situation occurs for other core-excited states, but the degree of the coupling to the
continuum and, therefore, the magnitude of instabilities differ for different states. In
particular, the $1s2s2p\,^4\!P^o$ state is rather insensitive to the interaction with the
continuum. On the contrary, the $^2\!D$ state is very much so. Moreover, the latter couples not
only with the $l = 1$ continuum states (as is the case for the $S$ and $P$ states), but also to the
$l = 2$ ones. Table \ref{tab:checkgridCI} presents a comparison of the convergence of the CI
energies of the $1s2p^2\,^2\!D_{3/2}$ state for $Z = 6$ and $Z = 17$, as obtained by different
methods. The left part of the table contains results obtained with the standard one-electron basis,
whereas the right part displays results obtained with the balanced basis. We observe that the
proposed balancing of the spectrum of one-electron orbitals significantly improves the convergence
of the calculated CI energies, which entails an improvement of the estimated accuracy by up to two
orders of magnitude.

Let us now discuss how do we produce a balanced discrete representation of the one-electron
continuum spectrum. In our approach, the one-electron orbitals are taken from the finite basis set
representation of the Dirac Hamiltonian with the frozen-core Dirac-Fock potential, obtained by the
DKB $B$-spline method \cite{shabaev:04:DKB}. The $B$-splines are defined on a radial grid, whose
form outside of the nucleus is exponential,
$$
t_i = t_0 e^{A\,i/N}\,, \ \ i = 0\,\ldots N\,,
$$
where $A = \ln(t_{\rm max}/t_0)$, $t_{\rm max}$ is the radial cutoff parameter and $t_0$ is the
nuclear radius. In the present work, we introduce a continuous parameter $\gamma$ in the definition
of the radial grid,
$$
t_i = t_0 e^{A\,(i/N)^{\gamma}}\,.
$$
After that, the energies of the continuum states of the $B$-spline representation of the
one-electron Dirac Hamiltonian spectrum become functions of the parameter $\gamma$, $\vare_n \equiv
\vare_n(\gamma)$. By varying $\gamma$ (typically, by 10-20\% from the standard value $\gamma = 1$),
we were able to adjust the energy positions of the two nearest continuum states to be symmetrical
with respect to the reference-state energy. In practice, we repeated our CI calculations for
different values of the parameter $\gamma$, adjusting this parameter until the separation energies
of the two closest-lying continuum states from the reference state were equal. In these
calculations it was sufficient to restrict the basis by $L_{\rm max} = 1$ for the $S$ and $P$
states and  by $L_{\rm max} = 2$ for the $D$ states, since the higher-$L$ continuum states do not
cause any problems.

\section{QED effects}
\label{sec:QED}

In the present work we evaluated the leading QED effects to the energy levels by means of the model
QED operator approach \cite{shabaev:13:qedmod}. We used the implementation of this method in the
form of the QEDMOD package \cite{shabaev:14:qedmod}. Since the published version of the package had
a restriction of the nuclear charge $Z \ge 10$, we had to extend it to the lower values of $Z$. We
did this by performing numerical calculations of the one-loop self-energy matrix elements for the
$ns$, $np_j$, and $nd_j$ states for $Z = 3 - 9$ by the method described in
Ref.~\cite{yerokhin:05:se} (extending Tables I - IV of Ref.~\cite{shabaev:13:qedmod}).

In order to establish the level of accuracy of the model QED operator approach in the region of low
nuclear charges, we compared values obtained with the QEDMOD package with results of rigourous QED
calculations for the $1s^22s$ and $1s^22p_{1/2}$ states. Such calculations were accomplished for
the lightest Li-like atoms within the NRQED approach  \cite{puchalski:08:li} and for heavier
Li-like ions within the all-order in $\Za$ method \cite{yerokhin:06:prl,kozhedub:10}. The
comparison is presented in Table~\ref{tab:qed}. From this comparison, we conclude that the QEDMOD
package reproduces results of rigorous QED calculations for the total energies to accuracy of
better than 1\%.

\begin{table}
\caption{The QED correction to the total energy of the $1s^22s$ and $1s^22p_{1/2}$ states:
comparison of results of
rigourous QED calculations (``exact'') with ones obtained by the
QEDMOD package. The
``exact'' results for the ionization energies of Li-like ions are taken from
Ref.~\cite{puchalski:08:li} for Li and Be$^+$ and from Ref.~\cite{kozhedub:10} for Mg$^{9+}$ and Ar$^{15+}$.
In order to obtain QED shifts to the total energies, we added results from
Refs.~\cite{yerokhin:10:helike,artemyev:05:pra} for the ionization energy of the corresponding He-like ions and
results from Ref.~\cite{yerokhin:15:Hlike} for the ionization energy of H-like ions.
Units are a.u.
\label{tab:qed}
}
\begin{ruledtabular}
\begin{tabular}{lllll}
\multicolumn{1}{c}{$Z$}  &
      &  \multicolumn{1}{c}{$1s^22s$}   &  \multicolumn{1}{c}{$1s^22p_{1/2}$}
            & \multicolumn{1}{c}{$2s$-$2p_{1/2}$} \\
\hline\\[-5pt]
3 & exact     & 0.000\,114\,53\,(5)  &  0.000\,113\,14\,(7) &  0.000\,001\,388\,(4) \\
  & QEDMOD    & 0.000\,115\,3        &  0.000\,113\,9    &     0.000\,001\,42  \\
4 & exact     & 0.000\,350\,9\,(4)   &  0.000\,342\,1\,(6) &   0.000\,008\,83\,(4) \\
  & QEDMOD    & 0.000\,353\,         &  0.000\,345\,     &     0.000\,008\,45  \\
12& exact     & 0.020\,27            &  0.019\,21        &     0.001\,06  \\
  & QEDMOD    & 0.020\,38            &  0.019\,33        &     0.001\,05  \\
18& exact     & 0.084\,10            &  0.079\,29        &     0.004\,81  \\
  & QEDMOD    & 0.084\,42            &  0.079\,64        &     0.004\,78   \\
\end{tabular}
\end{ruledtabular}
\end{table}

\section{Results and discussion}

Numerical results of our calculations of the energy levels for the ground $1s^22s$ state, the first
two valence-excited $1s^22p_j$ states and the core-excited $1s2l2l'$ states of Li-like ions are
listed in Table~\ref{tab:en}. The results are given for ions along the lithium isoelectronic
sequence starting from carbon ($Z = 6$) and ending with chlorine ($Z = 17$). For the ground state,
the table presents ionization energies, i.e., energies relative to the ground state of the
corresponding He-like ion. For the valence-excited states, we present energies of the
$1s^22p_{1/2}$ state relative to the ground $1s^22s$ state and the $2p_{3/2}$-$2p_{1/2}$
fine-structure interval, since they usually appear in the literature in this form. For the
core-excited levels, the center-of-gravity (cg) energies relative to the ground state and the
fine-structure intervals ($J$-$J'$) are listed.

\newpage
\begin{longtable*}{llcdddddd}
\caption{Energy levels of Li-like ions, in Rydbergs. The labelling is as follows: ``Coul'' stands
for the Dirac-Coulomb energies, ``Breit'' denotes the correction due to the Breit interaction,
``NMS'' is the normal mass shift, ``SMS'' is the specific mass shift, ``QED'' denotes the QED
correction. Further quantities are: $\mu$,  the reduced mass, $R$, the root-mean-square nuclear
charge radius, and $\rm cg$, the center-of-gravity energy of the multiplet level. \label{tab:en}}
\\ \colrule\hline
\multicolumn{2}{c}{Term}  &
       \multicolumn{1}{c}{$J$}  &
                 \multicolumn{1}{c}{Coul}  &
                             \multicolumn{1}{c}{Breit}  &
                                           \multicolumn{1}{c}{NMS}  &
                                                         \multicolumn{1}{c}{SMS}  &
                                                                      \multicolumn{1}{c}{QED}  &
                                                                                           \multicolumn{1}{c}{Total}\\
\hline\\[-15pt]            %
%
%
%
%
\\ \multicolumn{6}{l}{
{\bf C  (Z =   6)},\ \  $1-\mu/m = 0.00004573$,\ \ $R = 2.470$\ fm} \\[2pt]
$1s^22s$                   & $^2\!S$   & 1/2       &        -4.74074 &    0.00022 &    0.00022 &    0.00001 &    0.00011 &      -4.74019\,(10) \\
$1s^22p$                   & $^2\!P^o$ & 1/2       &         0.58774 &    0.00013 &   -0.00003 &   -0.00011 &   -0.00012 &       0.58761\,(11) \\
             &  & 3/2-1/2                          &         0.00118 &   -0.00023 &    0.0     &    0.0     &    0.0     &        0.00096\,(8) \\
\end{longtable*}

Table~\ref{tab:en} presents the total theoretical energies as well as individual theoretical
contributions. For each level the Dirac-Coulomb energy and corrections to it due to the Breit interaction, the
normal mass shift (NMS), the specific mass shift (SMS), and the QED effects are provided. The theoretical
uncertainty comes from two main sources: the Dirac-Coulomb-Breit (DCB) energy and the QED
correction. The uncertainty of the DCB energies was estimated by performing a series of CI
calculations with 20-30 different basis sets and by analysing consecutive increments of the results
as the basis set was increased in different directions (see Ref.~\cite{yerokhin:12:lilike} for
details). The uncertainty of the QED energy shifts was estimated on the basis of the analysis presented
in Sec.~\ref{sec:QED}. For single energy levels, we assume the uncertainty of the QED correction to
be 1\%. For the energy differences, we take the smallest of the two values, 4\% of the QED
correction to the energy difference and the two QED uncertainties for the two states added
quadratically.

For the fine-structure intervals, the QED correction and its uncertainty largely cancel in the
difference, so our theoretical values are more accurate for these intervals than for the
center-of-gravity energies. The remaining theoretical uncertainty is dominated by the estimated
error of the DCB energies, obtained by adding quadratically the uncertainties for the two levels.

Our final theoretical results for the wavelengths of the 22 strongest $1s2l2l'$$\to$$1s^22l$
transitions are summarized in Table~\ref{tab:wav}. The transitions are labelled from ``$a$'' to
``$v$'', following the widely used notations by Gabriel~\cite{gabriel:72}.

%
%
%
\begingroup
\begin{ruledtabular}
\begin{longtable*}{cllllllllllllll}
\caption{Theoretical wavelengths of the $1s 2l2l'$$\to$$1s^2 2l$ transitions in
  lithlium-like ions, in \AA. Transitions are labelled as by Gabriel \cite{gabriel:72}.
\label{tab:wav}}\\
\colrule\hline\\[-6pt]
 Key & \multicolumn{1}{c}{Transition} & \multicolumn{1}{c}{$Z = 6$} & \multicolumn{1}{c}{$Z = 7$} & \multicolumn{1}{c}{$Z = 8$} & \multicolumn{1}{c}{$Z = 9$} & \multicolumn{1}{c}{$Z = 10$} & \multicolumn{1}{c}{$Z = 11$}\\
 \hline\\[-6pt]
$a$ & $1s(^2\!S)2p^2(^3\!P)\,^2\!P_{3/2} \to 1s^22p\,^2\!P^o_{3/2}$    & 41.36411\,(27)& 29.46580\,(15)& 22.04950\,(10) & 17.117058\,(77)& 13.671412\,(60)& 11.169775\,(52)\\
$b$ & $1s(^2\!S)2p^2(^3\!P)\,^2\!P_{3/2} \to 1s^22p\,^2\!P^o_{1/2}$    & 41.36230\,(26)& 29.46357\,(15)& 22.04692\,(10) & 17.114208\,(75)& 13.668325\,(60)& 11.166499\,(51)\\
$c$ & $1s(^2\!S)2p^2(^3\!P)\,^2\!P_{1/2} \to 1s^22p\,^2\!P^o_{3/2}$    & 41.36661\,(25)& 29.46851\,(15)& 22.05234\,(10) & 17.120020\,(76)& 13.674489\,(60)& 11.172962\,(52)\\
$d$ & $1s(^2\!S)2p^2(^3\!P)\,^2\!P_{1/2} \to 1s^22p\,^2\!P^o_{1/2}$    & 41.36480\,(25)& 29.46628\,(15)& 22.04977\,(10) & 17.117169\,(74)& 13.671401\,(60)& 11.169684\,(51)\\
\hline\hline
\end{longtable*}
\end{ruledtabular}
\endgroup

\begin{table}
\caption{Ionization potential of the ground state of Li-like ions, in eV.  \label{tab:io}}
\begin{ruledtabular}
\begin{tabular}{llll}
\multicolumn{1}{c}{$Z$}   &  \multicolumn{1}{c}{This work}   &  \multicolumn{1}{c}{Rigorous QED }
&  \multicolumn{1}{c}{NIST} \\
   &    &  \multicolumn{1}{c}{\cite{yerokhin:00:prl,yerokhin:01:2ph,yerokhin:07:lilike,kozhedub:10}}
&  \multicolumn{1}{c}{\cite{nist}} \\
\hline\\[-5pt]
10 & 239.097\,(2)  & 239.097\,(1) & 239.097\,(2) \\
12 & 367.501\,(2)  & 367.500\,(1) & 367.489\,(7) \\
15 & 611.755\,(3)  & 611.752\,(1) & 611.741\,(7) \\
\end{tabular}
\end{ruledtabular}
\end{table}

\begin{table}
\caption{Comparison of theory and experiment for the $2p_{1/2}$-$2s$ transition energy, in a.u. \label{tab:2p2s}}
\begin{ruledtabular}
\begin{tabular}{llll}
\multicolumn{1}{c}{$Z$}   &  \multicolumn{1}{c}{This work}   &  \multicolumn{1}{c}{Other theory}
            & \multicolumn{1}{c}{Experiment} \\
\hline\\[-5pt]
6 & 0.29380\,(5)    &         &           0.293811\,(5) $^a$ \\
  &                 &         &           0.293823\,(7) $^b$ \\
7 & 0.36660\,(6)    & 0.36663$^c$ &       0.366616\,(5) $^a$ \\
8 & 0.43909\,(6)    & 0.43911$^c$ &       0.439117\,(5) $^a$ \\
9 & 0.51147\,(6)    & 0.51150$^c$ &       0.51150\,(1)  $^a$ \\
10& 0.58385\,(7)    & 0.58387$^c$ &       0.58390\,(1)  $^a$  \\
  &                 & 0.58388\,(1)$^d$ &                    \\
11& 0.65637\,(9)    & 0.65637$^c$ &       0.65640\,(1)  $^a$  \\
12& 0.72903\,(9)    &         &           0.72907\,(1)  $^a$  \\
13& 0.80189\,(12)   & 0.80190$^c$ &       0.80197\,(1)  $^a$  \\
14& 0.87502\,(13)   &         &           0.87509\,(1)  $^a$  \\
15& 0.94847\,(14)   & 0.94846$^c$ &       0.94849\,(5)  $^a$  \\
  &                 & 0.94854\,(1)$^d$ &                    \\
16& 1.02224\,(16)   &         &           1.02232\,(3)  $^a$  \\
17& 1.09640\,(19)   & 1.09638$^c$ &       1.09652\,(5)  $^a$ \\
\end{tabular}
\end{ruledtabular}
$^a\,$ Ref.~\cite{edlen:83},\\
$^b\,$ Ref.~\cite{schramm:05},\\
$^c\,$ MBPT results from Ref.~\cite{johnson:88:b} and QED results from Ref.~\cite{mckenzie:91},\\
$^d\,$ Ref.~\cite{kozhedub:10}.
\end{table}

\begin{table}
\caption{Comparison of theory and experiment for the fine-structure splitting intervals of the $1s^22p\,^2\!P$,
 $1s2s2p\,^4\!P$, and
$1s2p^2\,^4\!P$ states, in cm$^{-1}$. \label{tab:fs}}
\begin{ruledtabular}
\begin{tabular}{llc..c}
\multicolumn{1}{c}{$Z$}
&  State
        &  $J-J'$
            &  \multicolumn{1}{c}{This work}
            & \multicolumn{1}{c}{Experiment}
            & \multicolumn{1}{c}{Ref.}\\
\hline\\[-5pt]
6 & $1s^22p\,^2\!P$    & 3/2-1/2 &  106x\,(8) &  107x.3    & \cite{edlen:83} \\
  & $1s2s2p\,^4\!P^o$  & 5/2-3/2 &   97x\,(1) &  100x\,(5) & \cite{livingston:78} \\
  &                    &         &            &   96x\,(2)&  \cite{engstrom:87} \\
  &                    & 3/2-1/2 &    4x      &    0x\,(7) & \cite{livingston:78} \\
  &                    &         &            &    4x\,(3)&  \cite{engstrom:87} \\
  & $1s2p^2\,^4\!P^e$  & 5/2-3/2 &   40x\,(2) &   41x\,(5) & \cite{livingston:78} \\
\end{tabular}
\end{ruledtabular}
\end{table}

\begin{table}
\caption{Comparison of theory and experiment for the center-of-gravity
multiplet separation wavelengths, in \AA. \label{tab:separation}}
\begin{ruledtabular}
\begin{tabular}{lddc}
\multicolumn{1}{c}{$Z$}
            &  \multicolumn{1}{c}{This work}
            & \multicolumn{1}{c}{Experiment}
            & \multicolumn{1}{c}{Ref.}\\
\hline\\[-5pt]
\multicolumn{4}{l}{$1s2p^2\,^4\!P^e$ -- $1s2s2p\,^4\!P^o$}\\
6  &     1344.41\,(21) &   1344.2\,(3)   &  \cite{livingston:78} \\
   &                   &   1344.49\,(4)  &  \cite{engstrom:87} \\
7  &     1110.94\,(15) &   1111.1\,(1)   &  \cite{livingston:78} \\
8  &      945.87\,(11) &    946.8\,(1)   &  \cite{livingston:78} \\
9  &      822.54\,(9)  &    822.47\,(8)  &  \cite{martinson:83} \\
10 &      726.62\,(8)  &    726.51\,(5)  &  \cite{livingston:84} \\
   &                   &    726.71\,(20) &  \cite{knystautas:85} \\
12 &      586.24\,(8)  &    586.04\,(5)  &  \cite{traubert:82} \\
13 &      533.03\,(7)  &    532.83\,(9)  &  \cite{traubert:83} \\
   &                   &    533.11\,(9)  &  \cite{buchet:83}  \\[0.25cm]
\multicolumn{4}{l}{$1s2s2p\,(^1\!P^o) ^2\!P^o $-- $ 1s2s2p\,(^3\!P^o) ^2\!P^o$}\\
6  &    3592.(6.)      &   3582.\,(4.)   & \cite{mueller:09} \\
\end{tabular}
\end{ruledtabular}
\end{table}

\begin{table}
\caption{Comparison of theory and experiment for the center-of-gravity level energies, relative to the
ground $(1s)^2\,2s$ state, in eV. \label{tab:cog}}
\begin{ruledtabular}
\begin{tabular}{llddl}
\multicolumn{1}{c}{$Z$}
      & \multicolumn{1}{c}{State}
            &  \multicolumn{1}{c}{This work}
            & \multicolumn{1}{c}{Experiment}
            & \multicolumn{1}{c}{Ref.}\\
\hline\\[-5pt]
6  & $1s2s^2 {}^2\!S$                   &  291.597\,(8) & 291.59\,(10)  & \cite{mannervik:97}$^{\dag a}$ \\
   &                                    &               & 291.49\,(10) & \cite{mack:87}$^a$ \\
   & $1s2s2p\,(^3\!P^o) ^2\!P^o$            &  299.966\,(1) & 299.98\,(3)  & \cite{mueller:09} \\
   &                                    &               & 299.98\,(5)  & \cite{mannervik:97}$^{\dag a}$ \\
   &                                    &               & 299.91\,(7)  & \cite{nicolosi:77} \\
   & $1s2s2p\,(^1\!P^o) ^2\!P^o$            &  303.418\,(5) & 303.44\,(3)  & \cite{mueller:09} \\
\end{tabular}
\end{ruledtabular}
$^{\dag}$ the statistical uncertainty as given in Table~IV of Ref.~\cite{mannervik:97} is added
quadratically to the energy scale uncertainty of 0.05~eV mentioned in the text.\\
$^a$ using the ionization energy of the ground state of the corresponding ion from Table~\ref{tab:en}.
\end{table}

We now turn to analyzing the obtained results. We start with comparing our predictions with
benchmark theoretical and experimental results available for the ground and first excited states of
Li-like ions. Rigorous QED calculations to all orders in the nuclear binding strength parameter
$\Za$ were performed in
Refs.~\cite{yerokhin:00:prl,yerokhin:01:2ph,yerokhin:07:lilike,kozhedub:10}, yielding the presently
best theoretical results for the $1s^22s$, $1s^22p_{1/2}$, and $1s^22p_{3/2}$ states of Li-like
ions with $Z \ge 10$. The ionization potential of the $1s^22s$ state was not presented explicitly
but can be extracted. The corresponding comparison is presented in Table~\ref{tab:io}. Excellent
agreement  with the results of the rigorous QED calculations confirms that we were able to keep the
electron-correlation and QED effects in our calculations well under control.

Table~\ref{tab:2p2s} presents a comparison of our predictions for the $2p_{1/2}$-$2s$ transition
energies with the best theoretical and experimental data. There are remarkably many experimental
results with accuracy significantly better than that of our predictions, most of them produced
decades ago and summarized in Ref.~\cite{edlen:83}. The rigorous QED calculations
\cite{kozhedub:10} provide results only for $Z = 10$ and $15$ in the region of $Z$ relevant for the
present work. Because of this, we also compare our values against the MBPT results of
Ref.~\cite{johnson:88:b} supplemented by the QED correction evaluated separately in
Ref.~\cite{mckenzie:91}. Although these results do not have estimations of uncertainties, they
reproduce the experimental data remarkably well. We observe  agreement within the estimated error
bars for all cases listed in the table.

Table~\ref{tab:fs} compares our theoretical results for the fine-structure splitting intervals of
the $1s^22p\,^2\!P$,  $1s2s2p\,^4\!P^o$, and $1s2p^2\,^4\!P^e$ states with the available
experimental data. We observe that for the $1s^22p\,^2\!P$ state, the fine structure has been
measured up to an accuracy significantly higher than that of our theoretical predictions. The
agreement between theory and experiment is very good for $Z \le 13$, but for $Z  = 14$-$17$ we
observe deviations on the level of 2-3$\,\sigma$. For the $1s2s2p\,^4\!P^o$ and $1s2p^2\,^4\!P^e$
fine structure, our theory is more accurate than experiment.

In Table~\ref{tab:separation} we compare our theoretical results with the experimental data on the
multiplet separation center-of-gravity energies of the core-excited $P$ levels. In this case,
theory and experiment are on a similar level of accuracy. The agreement is very good for $Z \leq 10$
and deteriorates somewhat for $Z = 12$ and $13$.

Finally, in Table~\ref{tab:cog} we present a comparison of theoretical and experimental results for
the center-of-gravity energies of different levels relative to the ground state. In this case the
accuracy of our theory is significantly higher than that of the experimental data. The agreement is
good in the case of carbon. For higher-$Z$ ions, however, we observe numerous deviations between
theory and experiment, and between different experiments. We attribute these discrepancies to
difficulties to reliably assign the energy scale in such X-ray measurements.

\section*{Conclusion}

We performed extensive relativistic calculations of the energy levels and the fine-structure
splitting of the $n = 2$ valence and core-excited states of Li-like ions. The Dirac-Coulomb-Breit
energies were obtained by the configuration-interaction method adapted for treatment of
autoionizing core-excited states. By using specially balanced basis sets, we were able to improve
the convergence of our results and enhance the numerical accuracy of the calculated energy levels
by up to two orders of magnitude as compared to that for the standard basis. The uncertainty of the
Dirac-Coulomb-Breit energies was evaluated by analysing the convergence of the results with respect
to the number of partial waves included and the size of the one-electron basis.

The relativistic energies were supplemented with the QED energy shifts. To this end, the model QED
operator approach as implemented by the QEDMOD package was used. In order to cover the range of the
nuclear charge numbers $Z$ aimed at in the present work, we extended the QEDMOD package (originally
limited to $Z\ge 10$) to the lower values of $Z$. The uncertainty estimation  of the QED energy
shifts was based on the comparison with results of rigorous QED calculations available for the
ground and first excited states of Li-like ions.

The main result of the present work is the tabulation of theoretical energy levels and transition
wavelengths for the $1s^22l$ and $1s2l2l'$ states of ions along the lithium isoelectronic sequence
from carbon to chlorine. All our theoretical predictions are supplied with uncertainties that
include estimations of effects omitted in the theoretical treatment. For the $1s^22l$ states, our
results agree well with the benchmark theoretical and experimental results available in the
literature. For the core-excited $1s2l2l'$ energy levels, our theory is by an order of magnitude
more accurate than the measurements performed so far, which opens possibilities for using
theoretical predictions for calibrating experimental X-ray and electron spectra.

\section*{Acknowledgement}

V.A.Y. acknowledges support by the Ministry of Education and Science of the Russian Federation
Grant No.~3.5397.2017/6.7.


\end{document}